\documentclass[superscriptaddress,twocolumn]{jpsj3}

\usepackage{amsmath}
\usepackage{amsfonts}
\usepackage{amssymb}
\usepackage{txfonts}
\usepackage{bm}
\usepackage{tabularx}
\usepackage{graphicx,color}
\usepackage{ulem}
\usepackage{mathrsfs}
\def\Vec#1{\bm{#1}}
\def\Hc2{H_\mathrm{c2}}
\def\Tc{T_\mathrm{c}}

\begin{document}

\title{
Nodal Superconductivity of UTe$_2$ Probed by Field-Angle-Resolved Specific Heat \\on a Crystal with $\Tc=2.1$~K
}

\author{
	Kaito \textsc{Totsuka}$^{1,2}$,
	Yohei \textsc{Kono}$^{2}$, 
	Yusei \textsc{Shimizu}$^{3,4}$, 
	Ai \textsc{Nakamura}$^{4}$, 
	Atsushi \textsc{Miyake}$^{4}$, \\
	Dai \textsc{Aoki}$^{4}$, 
	Yasumasa \textsc{Tsutsumi}$^{5}$, 
	Kazushige \textsc{Machida}$^{6}$, and
	Shunichiro \textsc{Kittaka}$^{1,2}$\thanks{kittaka@g.ecc.u-tokyo.ac.jp}
}

\inst{$^{1}$Department of Basic Science, The University of Tokyo, Meguro, Tokyo 153-8902, Japan\\
      $^{2}$Department of Physics, Faculty of Science and Engineering, Chuo University, Kasuga, Bunkyo-ku, Tokyo 112-8551, Japan\\
      $^{3}$Institute for Solid State Physics, The University of Tokyo, Kashiwa, Chiba 277-8581, Japan\\
      $^{4}$IMR, Tohoku University, Oarai, Ibaraki 311-1313, Japan\\
      $^{5}$Department of Civil and Environmental Engineering, Faculty of Engineering, Hiroshima Institute of Technology, Hiroshima 731-5193, Japan\\
      $^{6}$Department of Physics, Ritsumeikan University, Kusatsu, Shiga 525-8577, Japan
}
\date{\today}

\abst{
Field-angle-resolved specific-heat measurements were performed on 
a clean single crystal of a spin-triplet superconductor UTe$_2$ with $\Tc=2.1$~K and a low residual electronic specific heat.
At low temperatures, the specific heat exhibits a linear dependence on the magnetic field when the field is applied precisely along the $b$ axis, 
in stark contrast to its rapid increase at low fields for other orientations.
This pronounced anisotropy suggests the presence of nodal quasiparticle excitations with the Fermi velocity predominantly aligned along the $b$ axis.
Considering the characteristic field-angle dependences of both the specific heat and the upper critical field, these observations are broadly compatible with theoretical models that assume a superconducting gap structure
featuring either point nodes consistent with $B_{\rm 2u}$ symmetry, allowed in the infinitely strong spin-orbit coupling scheme, or 
line nodes confined to flat regions of the quasi-two-dimensional Fermi surface, consistent with
 $^3B_{\rm 3u}$ symmetry in the finite spin-orbit classification scheme.
These results yield crucial hints for resolving the pairing symmetry of UTe$_2$, paving the way for a deeper understanding of its spin-triplet superconductivity.
}

\maketitle
\section{Introduction}
Spin-triplet superconductivity, distinguished by its rich internal degrees of freedom and potential for hosting unconventional quantum states, stands at the forefront of contemporary condensed matter physics.
Despite decades of theoretical predictions and experimental efforts, the identification of candidate materials has remained a formidable challenge, 
leaving the fundamental nature of this exotic pairing largely unresolved. 
The discovery of superconductivity in UTe$_2$ marked a turning point in this quest~\cite{Ran2019Science,Aoki2019JPSJ,Aoki2022JPCM}.
With its extraordinary superconducting properties, including a high upper critical field far exceeding the Pauli limit, 
field-reentrant superconductivity above 40~T~\cite{Ran2019NatPhys}, and multiple superconducting phases induced by magnetic field and pressure~\cite{Aoki2020JPSJ}, 
UTe$_2$ has rapidly emerged as a premier platform for investigating spin-triplet pairing.

UTe$_2$ crystallizes in the orthorhombic space group $Immm$ with the $a$ axis being the magnetization easy axis. 
Shortly after its discovery, the superconducting transition temperature $T_{\rm c}$ was reported to be 1.6~K. 
However, recent advances in crystal growth 
techniques have led to the synthesis of next-generation single crystals, 
in which $T_{\rm c}$ is significantly enhanced to 2.1~K~\cite{Sakai2022PRM,Aoki2024JPSJ2}. 
These high-quality crystals, characterized by an exceptionally low residual electronic specific-heat coefficient in the superconducting state, 
enable precise investigations of the intrinsic properties of UTe$_2$ without the complications arising from impurity and inhomogeneity. 
While the spin component has been strongly suggested to be of triplet character, based on comprehensive NMR studies~\cite{Nakamine2019JPSJ,Fujibayashi2022JPSJ,Nakamine2021PRB,Kinjo2023SA,Matsumura2023JPSJ},
the orbital part of the pairing function, namely, the superconducting gap structure, remains controversial.
Various experimental techniques, including thermal conductivity $\kappa$~\cite{Suetsugu2024SA,Hayes2025PRX}, magnetic penetration depth~\cite{Ishihara2023NC}, quasiparticle interference~\cite{Wang2025NatPhys}, and specific heat~\cite{Lee2025PRR},
have yielded seemingly conflicting evidence for both fully-gapped and nodal superconductivity.
In addition, while magnetic penetration depth measurements suggest a multi-component order parameter~\cite{Ishihara2023NC}, 
Kerr-effect and ultrasound studies point toward a single-component scenario in the orbital part of the pairing function~\cite{Ajeesh2023PRX,Theuss2024NP}.
Furthermore, the geometry of the Fermi surface (FS), whether predominantly two-dimensional or three-dimensional, adds another layer of complexity to the debate~\cite{Jiao2019,Fujimori2019JPSJ,Miao2020PRL,Aoki2022JPSJ,Eaton2024NC}.

To address these unresolved issues, particularly the nodal structure of the superconducting gap, 
we perform low-temperature specific-heat measurements on a clean single crystal of UTe$_2$, 
with precise control of the magnetic-field orientation.
Our results provide thermodynamic evidence for low-energy excitations of nodal quasiparticles
whose Fermi velocity is oriented along the $b$ axis.

\section{Methods}
High-quality single crystals were grown using the molten salt flux liquid transport method~\cite{Aoki2024JPSJ2}.
For this study, a thin, rectangular single crystal with a mass of 11.68~mg, elongated along the $a$ axis, was selected.
The surface of the sample parallel to the $ab$ plane was securely attached to the calorimeter stage using GE varnish.
Subsequently, the calorimeter was mounted on a dilution refrigerator (Oxford, Kelvinox AST Minisorb) 
such that the $c$ axis of the sample was aligned with the vertical $z$ direction.
This refrigerator was inserted into a vector magnet system~\cite{Sakakibara2016RPP}, 
which can generate magnetic fields up to 7~T (3~T) along the horizontal $x$ ($z$) direction.
The refrigerator can be rotated around the $z$ axis using a stepper motor.
Using this system, %together with measurements of the field-angle dependence of the specific heat in UTe$_2$, 
we achieved three-dimensional control of the magnetic-field orientation with an angular resolution better than $0.1^\circ$.
The heat capacity was measured using the quasi-adiabatic heat-pulse method.
The addenda heat capacity was measured separately and exhibited a Schottky-type anomaly in its field dependence at low temperatures~\cite{Kittaka2018JPSJ}.
This addenda contribution has been subtracted from all data presented below.

\section{Results and Discussion}
\subsection{Temperature dependence of the specific heat}
Figure~\ref{fig1}(a) shows the temperature dependence of the specific heat divided by temperature, $C/T$, 
where the phonon contribution has been subtracted. 
In zero field, a sharp specific-heat jump, $\Delta C/(\gamma T_{\rm c})=2.5$, is observed at $\Tc=2.05$~K with a transition width of about 0.06~K (onset at 2.1~K),
where $\gamma=125$~mJ mol$^{-1}$ K$^{-2}$ is the electronic specific heat coefficient in the normal state.
These results, together with the small value of $C/(\gamma T)=0.06$ at $0.1T_{\rm c}$, demonstrate the high quality of the present sample.

As exemplified by the data for $B \parallel c$ in Fig.~\ref{fig1}(a), the temperature dependence of $C/T$ under magnetic fields for each orientation generally resembles the behavior reported in Ref.~\ref{Lee2025PRR}, 
which investigated a sample with $T_{\rm c}=2.1$~K and $C/(\gamma T) \sim 0.12$ at $0.1T_{\rm c}$ in zero field.
In particular, our data reproduce the anomalous $T^2$ dependence of $C/T$ for $B \parallel a$ above 1~T, as reported in Supplementary Fig. S4 of Ref.~\ref{Lee2025PRR},
whose contribution becomes more pronounced with increasing temperature.
In addition, an upturn in the specific heat was observed below 0.15~K, similar to that previously reported for low-$T_{\rm c}$ samples~\cite{Kittaka2020PRR}, 
which is unlikely to originate from the nuclear contribution.
The origins of both the $T^2$ behavior and the low-temperature upturn remain unclear.
To minimize these contributions, field-angle-resolved specific-heat measurements in this study were primarily performed at 0.3~K ($\sim 0.15T_{\rm c}$).
At this temperature, the influence of anomalous components is reduced, enabling the detection of low-energy quasiparticle excitations around nodes,
as demonstrated in previous studies on various unconventional superconductors~\cite{Kittaka2012PRB,Kittaka2013JPSJ,Kittaka2014PRL,Shimizu2016PRL}. 

\subsection{Magnetic field dependence of the specific heat}
\begin{figure}
\includegraphics[width=3.4in]{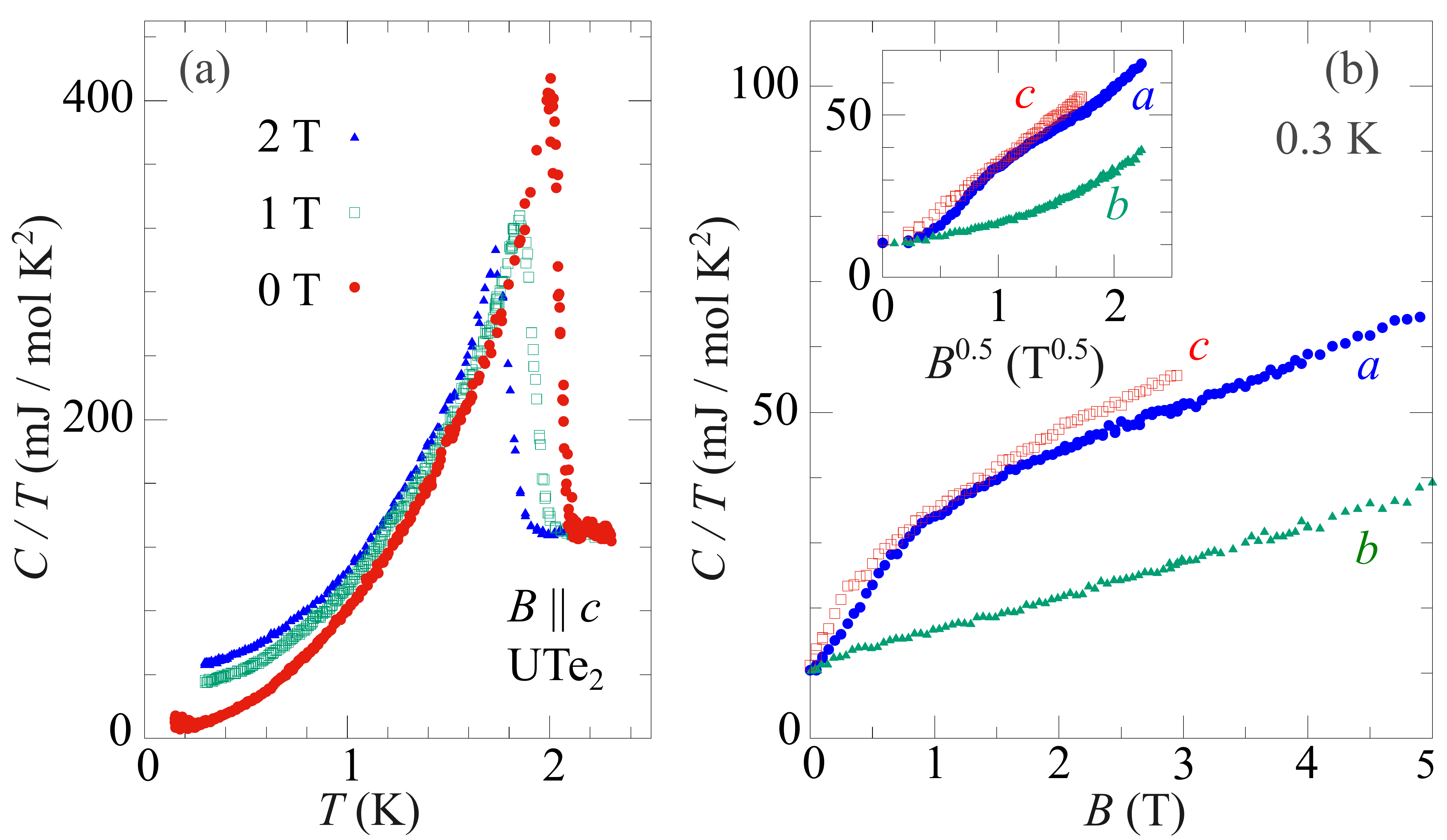}
\caption{
(Color online) 
(a)~Temperature dependence of $C/T$ under magnetic fields applied along the $c$ axis. 
(b)~Magnetic-field dependence of $C/T$ at 0.3~K for each field orientation. 
The inset shows the same data plotted as a function of $\sqrt{B}$.
}
\label{fig1}
\end{figure}

Figure \ref{fig1}(b) shows the magnetic-field dependence of the specific heat for fields applied along each axis at 0.3~K. 
When the magnetic field is applied along the $a$ or $c$ axis, 
a rapid increase in $C(B)$, reminiscent of the $\sqrt{B}$ ($B^{0.64}$) behavior expected for line-node (point-node) superconductors, is observed in the low-field region~\cite{Miranovic2005PRB},
as confirmed in the inset of Fig.~\ref{fig1}(b).
Although Ref.~\ref{Lee2025PRR} reported a $\sqrt{B}$-like dependence of the specific heat for all magnetic-field directions, 
our measurements, with fine tuning of the field orientation and small field increments, revealed that 
$C(B)$ is nearly proportional to $B$ along the $b$ axis.
The previous results obtained using a low-$T_{\rm c}$ sample with $C/(\gamma T) \sim 0.6$ at $0.1T_{\rm c}$ in zero field~\cite{Kittaka2020PRR} qualitatively differ from the present findings. 
This discrepancy is likely attributable to impurity effects in the earlier sample, which may have masked the signatures of low-energy quasiparticle excitations.

When the superconducting gap possesses nodes, low-energy quasiparticles are excited due to the Doppler energy shift given by $\Delta E=m_{\rm e}\Vec{v}_{\rm F}\cdot \Vec{v}_{\rm s}$,
where $m_{\rm e}$ is the electron mass, $\Vec{v}_{\rm F}$ is the Fermi velocity, and $\Vec{v}_{\rm s}$ is the local superfluid velocity circulating around vortex cores, 
which is perpendicular to the field direction.
This so-called Volovik effect~\cite{Volovik1993JETPL} leads to a $\sqrt{B}$-like behavior in the zero-energy quasiparticle density of states (ZDOS), $N(E=0)$, 
which can be detected through low-temperature specific-heat measurements.
However, when $\Vec{v}_{\rm F}$ at the nodes is uniquely oriented and the magnetic field is applied along this nodal $\Vec{v}_{\rm F}$ direction, i.e., $\Vec{v}_{\rm s}\perp\Vec{v}_{\rm F}$,
low-energy quasiparticle excitations are suppressed due to $\Delta E=0$.
In this case, a linear dependence $N(E=0) \propto B$ is expected, as it reflects the number of vortex cores.
Based on the simple framework described above, 
the results shown in Fig.~\ref{fig1}(b) provide firm thermodynamic evidence that
nodes exist on the FS, where $\Vec{v}_{\rm F}$ points along the $b$ axis, and are absent on regions with $\Vec{v}_{\rm F}$ pointing in other directions.

\subsection{Field-angle dependence of the specific heat}
In order to further investigate the gap structure, we have investigated the field-angle dependence of the specific heat.
Indeed, the field-angle dependence of $N(E=0)$ and/or $C$ for various types of nodal superconductors has been extensively studied from a theoretical perspective~\cite{Miranovic2005PRB,Vorontsov2006PRL,Vekhter2008PhysicaB,Hiragi2010JPSJ,Tsutsumi2016PRB}.
It should be noted that the FS geometry affects the specific-heat oscillations associated with gap anisotropy, 
including characteristic fields and temperatures where the oscillation sign reverses~\cite{Vekhter2008PhysicaB}.
For uniaxially symmetric point-node or line node gap structures on a three-dimensional FS, 
the maxima of the field-angle-dependent ZDOS $N(\theta)$ shift from the antinodal to the nodal direction 
with increasing magnetic field, accompanied by a reversal of the oscillation pattern between low and high fields~\cite{Tsutsumi2016PRB}.
Therefore, to reliably determine the nodal position and structure, 
it is crucial to analyze the field-angle dependence of the specific heat over a range of magnetic fields, 
taking into account the possible reversal of the oscillation pattern.

\begin{figure}
\includegraphics[width=3.3in]{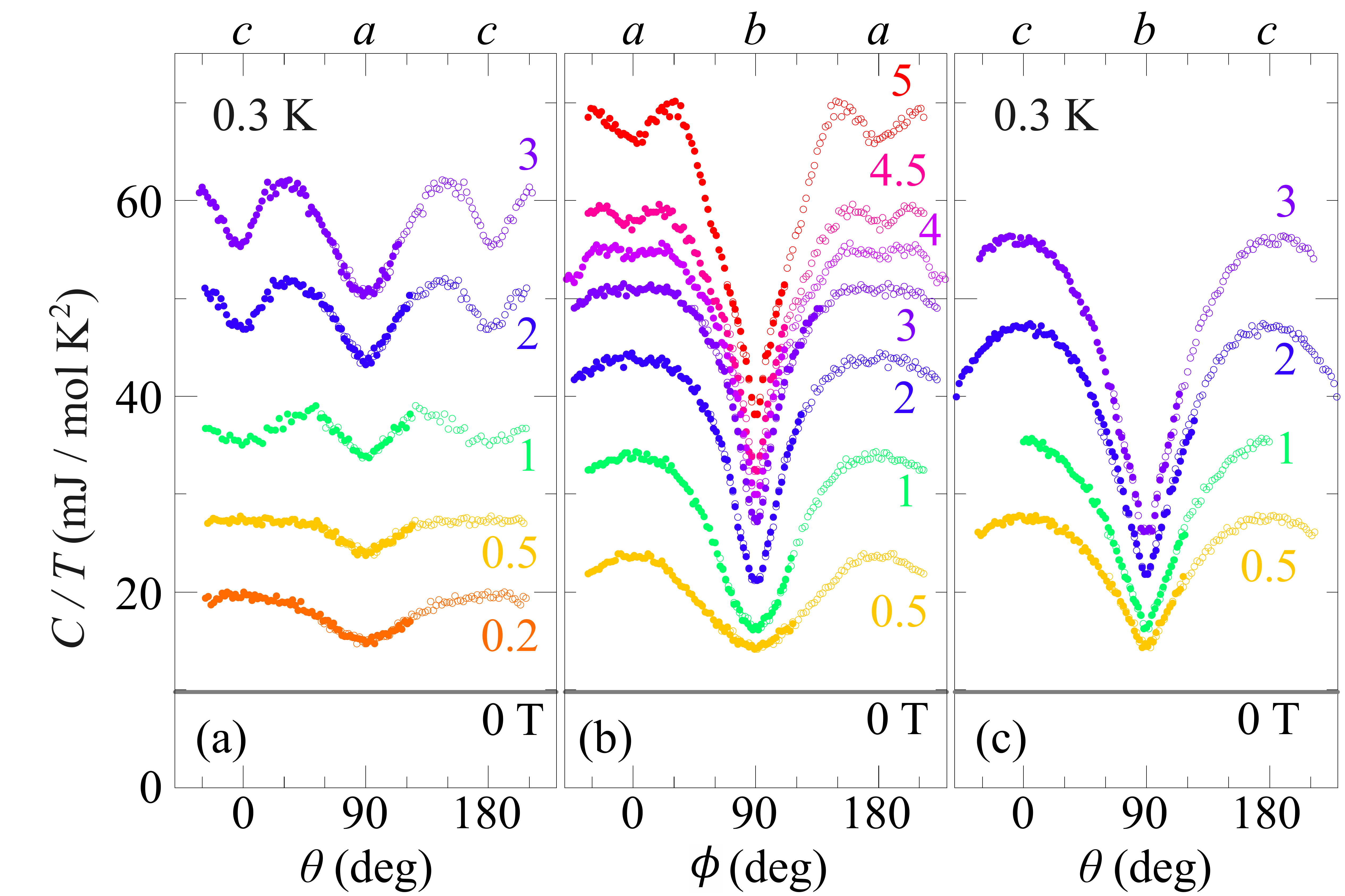}
\caption{
(Color online) 
Field-angle dependence of $C/T$ at 0.3~K under various magnetic fields rotated within the (a)~$ac$, (b)~$ab$, and (c) $bc$~planes. 
The numbers labeling each set of data indicate the magnetic-field strength in tesla.
Open symbols represent data points mirrored with respect to the crystal symmetry axes.
}
\label{fig2}
\end{figure}

Figures~\ref{fig2}(a)-\ref{fig2}(c) show the specific heat measured at 0.3~K under magnetic fields rotated within the $ac$, $ab$, and $bc$ planes, respectively. 
Here, the field angle $\theta$ ($\phi$) denotes the polar angle (azimuthal angle in the $ab$ plane) measured from the $c$ ($a$) axis. 
In each rotational plane, the specific heat exhibits a characteristic, field-dependent oscillation pattern, as described in detail below. 

\subsubsection{$ac$-plane field rotation}
We first focus on the oscillation pattern of the specific heat $C(\theta)$ obtained for field rotation within the $ac$ plane at $\phi=0^\circ$, as shown in Fig.~\ref{fig2}(a).
At low magnetic fields, $C(\theta)$ exhibits a minimum at $\theta=90^\circ$ ($B \parallel a$), 
which remains unchanged at higher fields.
Above 0.75~T, a dip develops at $\theta=0^\circ$ ($B\parallel c$), accompanied by a local maximum around $30^\circ \lesssim \theta \lesssim 45^\circ$.
Based on the $C(B)$ data in Fig.~\ref{fig1}(b), no nodes can exist within the $ac$ plane.
This suggests that nodal quasiparticles do not significantly contribute to the anisotropy in $C(\theta)$ at $\phi=0^\circ$. 
Therefore, the observed oscillation in Fig.~\ref{fig2}(a) is most reasonably attributed to the anisotropy of the upper critical field $B_{\rm c2}$, 
as $C(B)$ generally scales with $B/B_{\rm c2}$.

In this material, the anisotropy of $B_{\rm c2}$ within the $ac$ plane exhibits a pronounced temperature and field dependence~\cite{Aoki2024JPSJ}: near $T_{\rm c}$ and at low fields, $B_{\rm c2}^{\parallel a}>B_{\rm c2}^{\parallel c}$, whereas at low temperatures and high fields, the relation reverses to $B_{\rm c2}^{\parallel c}>B_{\rm c2}^{\parallel a}$. This behavior suggests a possible inversion of superconducting anisotropy with increasing field, although its microscopic origin remains unresolved. 
The observed $\theta$ dependence of the specific heat within the $ac$ plane can thus be interpreted as part of this crossover process. Indeed, Ref.~\ref{Lee2025PRR} reports a reversal of the specific heat anisotropy around 6~T. 
Therefore, the crossover in the oscillation pattern of $C(\theta)$ for $B \parallel ac$ with increasing field strength 
likely reflects a field-induced change in the superconducting anisotropy, rather than a change driven by nodal quasiparticle excitations.

\subsubsection{$ab$-plane field rotation}
We next turn to the results of $C(\phi)$ for $B \parallel ab$, shown in Fig.~\ref{fig2}(b).
Assuming that nodes exist only along the $b$ axis, quasiparticle excitations are nearly isotropic when the field is rotated within the $ac$ plane. In contrast, for rotations within planes that include the $b$ axis, the field direction relative to the node strongly affects quasiparticle excitations, making the influence of the gap structure more pronounced than in the $ac$ plane. 

Across the present field range, $C(\phi)$ at $\theta=90^\circ$ consistently exhibits a minimum along the $b$ axis.
Above 3~T, a dip develops around the $a$ axis ($\phi=0^\circ$); 
the maximum shifts from the $a$ toward the $b$ axis, accompanied by a shoulder or peak anomaly at intermediate field angles.
Unlike $B_{\rm c2}(\theta)$ in the $ac$ plane, $B_{\rm c2}(\phi)$ in the $ab$ plane does not show a prominent minimum at intermediate field angles~\cite{Aoki2024JPSJ}.
Therefore, the observed oscillations in $C(\phi)$ cannot be explained solely by the anisotropy of $B_{\rm c2}$.
Theoretical models assuming a spherical FS and the presence of nodes with $\Vec{v}_{\rm F} \parallel b$ predict a shoulder or peak anomaly 
in the angular dependence of $N(E=0)$ within the $ab$ plane, appearing at an intermediate field angle~\cite{Tsutsumi2016PRB,Shimizu2016PRL,Kittaka2020PRR}.
The consistency between experimental results and theoretical predictions supports the presence of nodes along $\Vec{v}_{\rm F} \parallel b$, as also indicated by the $C(B)$ data.

\subsubsection{$bc$-plane field rotation}
\begin{figure}
\includegraphics[width=3.3in]{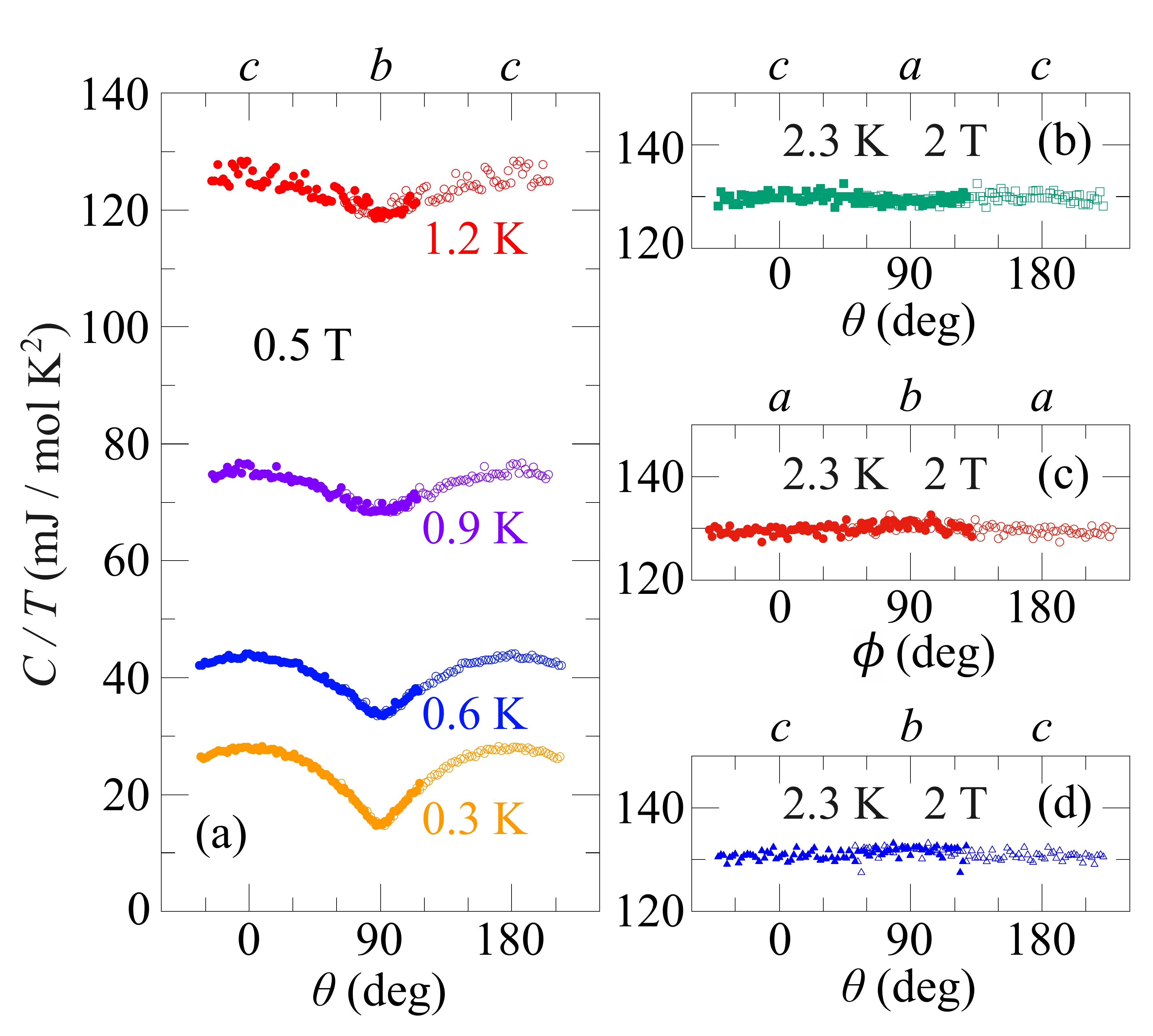}
\caption{
(Color online) 
(a) Field-angle dependence of $C/T$ in the superconducting state at various temperatures under a magnetic field of 0.5~T rotated within the $bc$ plane.
(b)-(d) $C/T$ in the normal state at 2.3 K under a magnetic field of 2 T rotated within the (b) $ac$, (c) $ab$, and (d) $bc$ planes.
Open symbols indicate data points mirrored with respect to the crystal symmetry axes.
}
\label{fig3}
\end{figure}

In contrast to the non-monotonic field-angle dependence of $C(\phi)$ for $B \parallel ab$, 
$C(\theta)$ at $\phi=90^\circ$ ($B\parallel bc$) [Fig. \ref{fig2}(c)] consistently exhibits a $|\cos\theta|$-like behavior with a minimum along the $b$ axis,
without showing a shoulder or peak anomaly.
Since the anisotropy of the coherence length at low fields is relatively small, 
this sharp-dip structure around $B \parallel b$ cannot be attributed to $B_{\rm c2}$ anisotropy,
but rather suggests low-energy quasiparticle excitations associated with point nodes or line nodes~\cite{Tsutsumi2016PRB,Izawa2002PRL}.
With increasing temperature, the sharp-dip feature in the $|\cos\theta|$-like behavior is gradually suppressed, as shown in Fig.~\ref{fig3}(a). 
This suppression may be attributed to thermal quasiparticle excitations around nodes where $\Vec{v}_{\rm F} \parallel b$, 
which becomes active upon warming even when $B \parallel b$.
In the normal state, as shown in Figs.~\ref{fig3}(b)-\ref{fig3}(d), no prominent field-angle dependence is observed in $C(\phi,\theta)$, 
in contrast to a previous report using a low-$T_{\rm c}$ sample~\cite{Kittaka2020PRR}.
This demonstrates that the low-temperature oscillation in $C(\phi,\theta)$ is an intrinsic feature of the superconducting state.

\subsection{Possible superconducting gap structures}
\subsubsection{Point-node scenario}
The absence of a shoulder or peak anomaly in $C(\theta)$ for $B \parallel bc$ is apparently incompatible 
with the theoretical prediction~\cite{Tsutsumi2016PRB} 
which assumes that quasiparticle excitations are dominated by point nodes located along $\Vec{v}_{\rm F} \parallel b$ 
on a three-dimensional FS.
This discrepancy may arise from the geometry of the FS.
If the dominant superconducting bands in UTe$_2$ are quasi-two-dimensional FSs extended along the $c$ axis, 
there would be many regions where $\Vec{v}_{\rm F} \parallel b$, whereas regions with $\Vec{v}_{\rm F} \parallel c$ would be absent.
It should be noted that a non-monotonic oscillation pattern, often accompanied by a shoulder or peak anomaly, can occur when the $N(E=0)$ or $\gamma(B) \equiv C(B)/T|_{T\rightarrow 0}$ curves for $B \parallel b$ and $B \parallel c$ approach each other and intersect.
As seen in Fig.~\ref{fig1}(b), the $C(B)$ curves for $B \parallel b$ and $B \parallel c$ show no sign of crossing, 
even at higher fields up to $B_{{\rm c2} \parallel c}\sim 14$~T~\cite{Lee2025PRR}.

\begin{figure}
\includegraphics[width=3.5in]{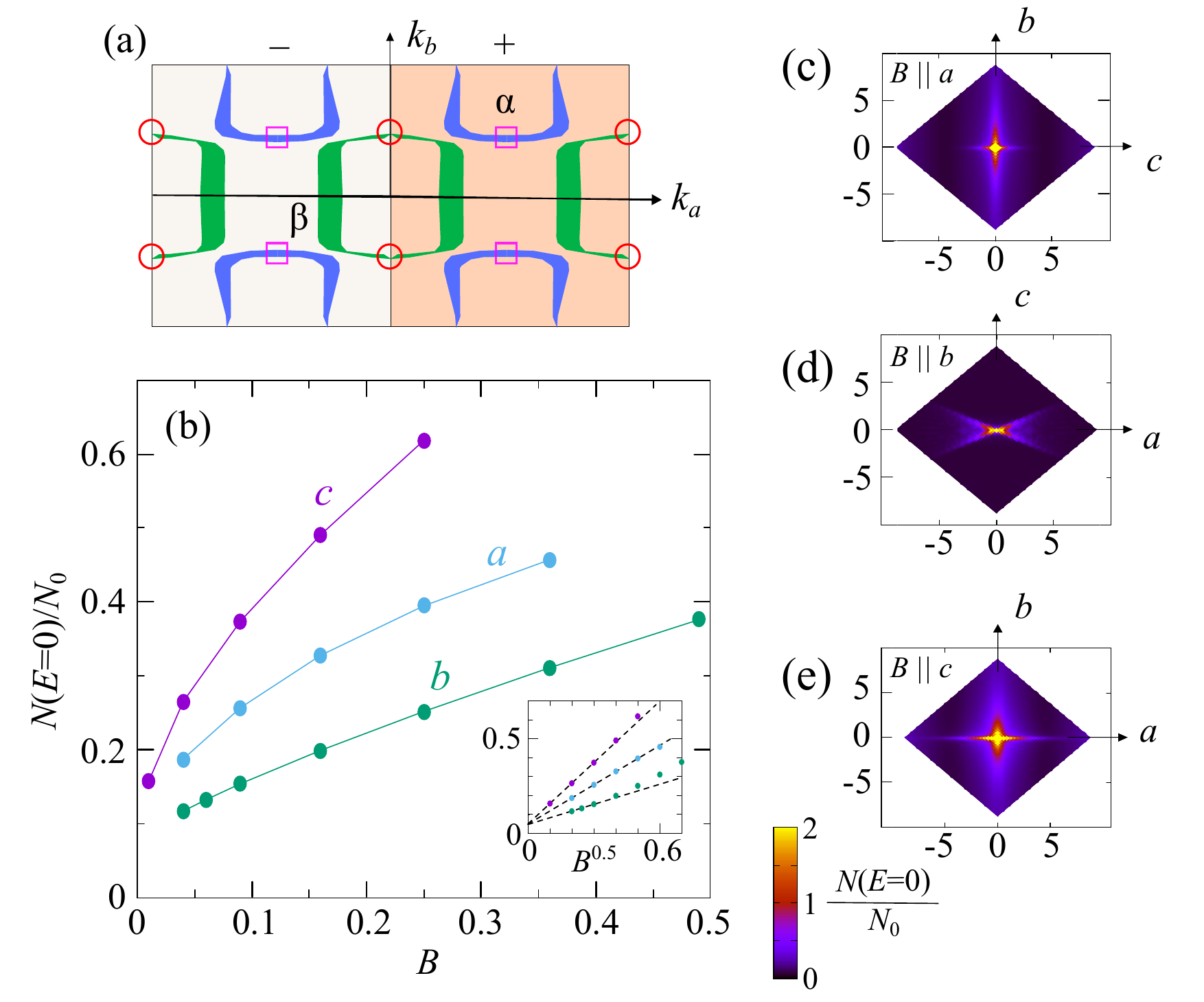}
\caption{
(Color online) 
(a) Top view of the 3D Fermi surfaces projected onto the $k_a$-$k_b$ plane. 
The thickness of the FS lines is roughly proportional to the warping along the $k_c$ axis. 
The red circles indicate the positions of the line nodes for ${^3}B_{\rm 3u}$. 
Point nodes are located at both circles and squares for $B_{\rm 2u}$.
The sign of the order parameter is alternate along the $k_a$ direction. 
The original figure is drawn
by Eaton \textit{et al}.~\cite{Eaton2024NC}.
(b) Calculated results of the ZDOS, $N(E=0)/N_0$ or $\gamma(B)$, 
as a function of $B$ for three field orientations.
The inset shows the same data plotted as a function of $\sqrt{B}$.
(c)-(e) The spatial profiles of the ZDOS centered at the vortex core in a unit cell for three field directions at $B=0.04$.
Here, $B$ is expressed in the Eilenberger unit~\cite{Tsutsumi2016PRB} and $T=0.2\Tc$.
}
\label{fig4}
\end{figure}

\subsubsection{Line-node scenario}
We now examine the alternative possibility of the line node scenario relative to the point node one
by backing up the microscopic quasi-classical Eilenberger theory~\cite{Ichioka1997PRB,Hayashi1997PRB} 
with a realistic FS model, consisting of the two cylindrical FSs~\cite{Eaton2024NC}
as shown in Fig.~\ref{fig4}(a). 
The FS information is critical in performing a realistic construction of the quasiparticle excitations in a vortex state to understand the angle-resolved specific heat data shown above.
It is not obvious to explain the linear field dependence of $C(B)\propto B$ for $B\parallel b$
because, when the magnetic field is applied along the nodal direction, the line nodes on the three dimensional FS 
necessarily have a component of $\Vec{v}_{\rm F}$ parallel to
$\Vec{v}_{\rm s}$.
This leads to the Doppler shift $\Delta E\propto\Vec{v}_{\rm F}\cdot\Vec{v}_{\rm s}$,
giving rise to the Volovik $\sqrt B$ effect in $\gamma(B)$ for $B \parallel b$.
However, quantum oscillation experiments
~\cite{Aoki2022JPSJ,Aoki2023JPSJ,Weinberger2024PRL,Eaton2024NC} and band calculations~\cite{Ishizuka2021PRB}
have revealed exceptionally flat FS regions on the $b$ face of the $\beta$ sheet and $a$ face of the $\alpha$ sheet.
The flatness of the $b$ face has also been confirmed by angular-dependent magnetoresistance oscillation measurements, specifically through the observation of the so-called Yamaji oscillations~\cite{Kimata2024}.
Eaton \textit{et al}.~\cite{Eaton2024NC} have constructed an analytical FS model (the Eaton model) consistent with these observations.

In our calculations, we adopt the Eaton model~\cite{Eaton2024NC} to examine the quasiparticle 
excitations and to narrow down the possible pairing symmetries realized in UTe$_2$. 
To best describe the experimental data discussed above, we assume a gap function $\Delta(\Vec{k})=\Delta_0\sin(ak_a/2)$
where $a$ is the lattice constant along the $a$ axis.
This gap structure has line nodes on the $b$ face
of the $\beta$ sheet, as indicated by circles in Fig.~\ref{fig4}(a).
In contrast, the $\alpha$ sheet remains fully gapped, although the gap function exhibits a sign change
within a unit cell, as illustrated by the different background colors labeled with $\pm$ in Fig.~\ref{fig4}(a). 

Figure~\ref{fig4}(b) presents the calculated field dependence of the ZDOS, $N(E=0)/N_0$, or equivalently $\gamma(B)$, 
for the three magnetic-field orientations. Here, $N_0$ denotes the density of states in the normal state.
It is evident that $\gamma(B)$ for $B \parallel c$ and $B \parallel a$ exhibit a $\sqrt{B}$ dependence, characteristic of the Volovik effect associated with  nodal quasiparticle excitations,
as highlighted in the inset where the data are plotted against $\sqrt B$.
In contrast, $\gamma(B)$ for $B \parallel b$ increases nearly linearly with $B$, indicating that no Doppler shift occurs
because $\Vec{v}_{\rm F}(k_c)$ at the nodal positions is perpendicular to $\Vec{v}_{\rm s}$ along the $k_c$ axis.
Therefore, the observed strict linear behavior $C(B)\propto B$ for $B\parallel b$, as shown in Fig.~\ref{fig1}(b), implies that
the FS hosting the line nodes must be flat, with no dispersion of $\Vec{v}_{\rm F}(k_c)$ along
the $k_c$ direction. 
The particular requirement is fulfilled by only the $b$ face of the $\beta$ sheet.
In such a case, the gap amplitude does not depend on the field angle within the $bc$ plane, where the magnetic field is rotated [see Fig.~\ref{fig2}(c)],
and no field-angle dependence of the specific heat arising from gap anisotropy would be expected.
Although the $a$ face of the $\alpha$ sheet also satisfies this requirement, as seen in Fig.~\ref{fig4}(a),
the experimental fact that $C(B)\propto \sqrt B$ for $B \parallel a$ excludes the possibility of line nodes along the $a$ axis.
We also note that the $b$ face of the $\alpha$ sheet has strong warping along the $k_c$ direction, 
making that the $\alpha$ sheet is not eligible for possessing line nodes.
Such warping would lead to a $\sqrt{B}$ dependence in $\gamma(B)$ for $B\parallel b$, which contradicts the observed linear behavior.

As shown in Fig.~\ref{fig4}(c) for $B\parallel a$ and Fig.~\ref{fig4}(e) for $B\parallel c$, the zero-energy quasiparticle spectral weight well extends toward the nodal directions along the $k_b$ axis.
In contrast, for $B\parallel b$ [Fig.~\ref{fig4}(d)], 
the spectral weight is confined near 
the vortex core at the center, still leaking out toward the off-axis directions. 
This results in a small $\sqrt B$ behavior barely seen in the lowest field region of Fig.~\ref{fig4}(b), 
although such behavior is not actually observed in Fig.~\ref{fig1}(b). 
The open angle of ZDOS is determined by the $\Vec{v}_{\rm F}$ anisotropy, namely $\tan^{-1}(\tilde{v}^c_{\rm F}/\tilde{v}^a_{\rm F})=\tan^{-1}(0.25/0.70)\sim 20^{\circ}$, where $\tilde{v}^c_{\rm F}$ denotes the root-mean-square average ${v}^c_{\rm F}$ over the FS evaluated in the Eaton model.
We notice that in recent STM experiments~\cite{STM1,STM2,STM3} performed for $B\parallel (0{\bar1}1)$
on the $(011)$ surface, which is tilted only by $\sim$24$^{\circ}$ 
from the $b$ axis, have revealed a similar ZDOS feature elongated along the $a$ direction and centered at the vortex core. 
 
The above calculated results uniquely identify the position of the line nodes described by $\Delta(\Vec{k})=\Delta_0\sin(ak_a/2)$.
Namely, the line nodes must reside on the $b$ face of the $\beta$ sheet. 
As shown in Fig.~\ref{fig4}(a), the gap function yields that (1) the line nodes are located on the $\beta$ sheet, and (2) the $\alpha$ sheet is fully gapped with a sign change.
This uniqueness arises from the particular FS structure, 
which has been determined through both experimental and theoretical investigations~\cite{Aoki2022JPSJ,Weinberger2024PRL,Eaton2024NC,Aoki2023JPSJ,Kimata2024,Ishizuka2021PRB}.
Away from the $k_b$ axis, the $\beta$ sheet begins to warp. 
Any nodal structures deviating from the $b$ axis would exhibit a $\sqrt B$ behavior for $B \parallel b$, which is incompatible with the strictly linear $B$ dependence observed in $C(B)$.

\subsection{Possible gap symmetries}
Let us discuss possible gap symmetries of UTe$_2$. The present specific-heat results suggest
the existence of nodes along the direction of $\Vec{v}_{\rm F} \parallel b$.
If the gap nodes are point-like, the only compatible single-component gap symmetry within the $D_{\rm 2h}$ point group, assuming an infinitely
strong spin-orbit coupling (SOC) scheme, is the $B_{\rm 2u}$ representation~\cite{Annett1990AP}. 
Both circles and squares in Fig.~\ref{fig4}(a) indicate the positions of the point nodes.
This symmetry is consistent with previous reports based on ultrasound and thermal-conductivity measurements~\cite{Theuss2024NP,Hayes2025PRX}.

Alternatively, a line-node gap symmetry is possible if one assumes a flat FS.
This type of line-node gap has not been widely discussed, primarily because it appears to violate the Blount theorem~\cite{Blount1985PRB}.
However, the theorem is strictly applicable only to classification schemes for triplet pairing in the limit of infinitely strong SOC.
According to the Blount theorem, symmetry-protected line nodes are forbidden under strong SOC,
where spin and orbital degrees of freedom are tightly coupled, except for non-symmorphic crystals~\cite{Norman2009PRB}, which is not the case for UTe$_2$. 
Conceptually, this apparent contradiction can be 
resolved by reconsidering the classification scheme: from the strong SOC limit, described by $D^{\rm spin+orbital}_{\rm 2h}\times U(1)$, to a finite SOC framework, described by ${\it SO}(3)^{\rm spin}\times D^{\rm orbital}_{\rm 2h}\times U(1)$~\cite{Annett1990AP,Ozaki1985PTP,Ozaki1986PTP}. 
In the strong SOC scheme, the Cooper pair moment is locked to the crystal lattice and cannot reorient under an applied magnetic field.
In contrast, the finite SOC scheme allows for $d$-vector rotation, consistent with experimental observations at $\sim 3$~T ($14$~T) for $B\parallel c$ ($B \parallel b$)~\cite{Nakamine2019JPSJ,Fujibayashi2022JPSJ,Nakamine2021PRB, Kinjo2023SA,Matsumura2023JPSJ}. 
Importantly, the finite SOC scheme generically permits the existence of line nodes~\cite{Annett1990AP,Ozaki1985PTP}.

The identified pairing function with line nodes along the $b$ axis belongs to
the ${^3}B_{\rm 3u}$ representation, namely $(\hat{b}+i\hat{c})\sin(ak_a/2)$ combined with the spin part~\cite{Annett1990AP},
and is consistent with the previously proposed non-unitary superconducting state~\cite{machida2020JPSJ,machida2024JLTP}.
This orbital one-component state is also compatible with the ultrasound measurements by Theuss \textit{et al}.~\cite{Theuss2024NP},
who exclude the orbitally two-component scenario, such as $\{B_{\rm 1u},A_{\rm u}\}$, 
due to the accidental degeneracy of two irreducible representations in the infinitely strong SOC scheme.

It is also important to examine the consistency of our results with other competing scenarios. 
Recent proposals of accidentally degenerate superconducting order parameters, such as the chiral $B_{\rm 3u} + iA_{\rm u}$ state~\cite{Ishihara2023NC} and the $B_{\rm 2u} + iB_{\rm 1u}$ or $B_{\rm 2u} + iA_{\rm u}$ states~\cite{Lee2025PRR}, allow for more complex nodal configurations. 
Our findings are consistent with these scenarios if multiple point nodes emerge only in regions where $v_{\rm F}$ is nearly parallel to the $b$ axis.
For example, previous penetration depth study supports the chiral $B_{\rm 3u} + iA_{\rm u}$ state in the zero-field limit, which possesses point nodes near the $k_y$ and $k_z$ axes~\cite{Ishihara2023NC}. This nodal structure is unlikely based on the present results. However, the application of a magnetic field can modify the chiral order parameter, potentially altering the positions of nodes under finite fields. As discussed by Ishihara et al.~\cite{Ishihara2023NC}, this possibility may resolve the apparent contradiction between our specific heat data and previous penetration depth study.

By contrast, recent thermal-conductivity measurements~\cite{Suetsugu2024SA} report a vanishingly small value of $\kappa/T$ and a lack of field dependence at low fields in the zero-temperature limit, 
which apparently contradict nodal superconductivity.
These results have been interpreted as evidence for a fully gapped superconducting state with $A_u$ symmetry.
One possible explanation for this discrepancy lies in 
thermal conductivity is primarily sensitive to light-mass quasiparticles, whereas specific heat captures contributions from the entire FSs, including heavy-mass quasiparticles. 
Moreover, thermal-conductivity measurements at low temperatures may be affected by electron-phonon decoupling, potentially leading to an underestimation of the electronic contribution~\cite{Behnia2025arXiv}.
Further studies are required to resolve the apparent discrepancy between the $b$-axis nodal scenario indicated by the present specific-heat measurements and the fully gapped state suggested by thermal conductivity data.

\section{Summary}
We have performed field-angle-resolved specific-heat measurements on a clean UTe$_2$ single crystal with $\Tc=2.1$~K and a low residual electronic specific heat. 
Our results suggest the occurrence of nodal quasiparticle excitations with $\Vec{v}_{\rm F} \parallel b$.
While the data are broadly consistent with a point-node gap of $B_{\rm 2u}$ symmetry, they also allow for the possibility of line nodes confined to flat regions of the hole Fermi surfaces, where $\Vec{v}_{\rm F}\parallel b$ holds across all $k_c$.
Our findings shed new light on the highly debated gap structure of UTe$_2$, offering evidence that helps elucidate its superconducting symmetry.

\section*{Acknowledgments}
We acknowledge T. Kanda for his valuable support. KM also thanks S. Lee and M. Kimata for discussion on their experiments.
This work was supported by JST FOREST Program (JPMJFR246O), JST SPRING (JPMJSP2108),  and JSPS KAKENHI (JP23H04868, JP23K25825, JP21K03455).

\vskip\baselineskip

\textit{Note added in proof---} 
After completing this manuscript, we noted a new preprint on field-angle-resolved heat transport in UTe$_2$ \cite{Hayes}, which provides a valuable complement to our study. 
We further draw the reader's attention to a recent theoretical report \cite{Machida2026JLTP}, which may offer useful insight into the present work.

\end{document}